\documentclass[10pt,twocolumn]{article}

\usepackage[english]{babel}
\usepackage[utf8x]{inputenc}
\usepackage[T1]{fontenc}
\usepackage{url}
\usepackage{hyperref}
\usepackage{float}

\hypersetup{hidelinks}

\usepackage[margin=1.25in,a4paper]{geometry}

\usepackage{amsmath}
\usepackage{graphicx}
\title{
		\usefont{OT1}{bch}{b}{n}
		\normalfont \normalsize \huge ACOUSTIC SCENE CLASSIFICATION \\USING AUDITORY DATASETS\\ 
}
\usepackage{authblk}
\author[1]{Jayesh Kumpawat}
\author[1]{Shubhajit Dey}
\affil[1]{INDIAN INSTITUTE OF SCIENCE EDUCATION AND RESEARCH, BHOPAL}

\begin{document}
\maketitle


\begin{abstract}

The approach of using audio datasets to solve the problems revolving around land use patterns has gained decent amount of research exposure in the recent times. Although being a blooming domain in the field of \textit{AI} and \textit{Data Analysis}, it would not be wrong to declare that a significant amount of journey is yet to be covered. In this article, the project conducted to classify some pre-defined acoustic scene is discussed and explained.\\

\centerline{\texttt{- \underline{System designed} -}}
\medskip

\begin{tabular}[t]{l p{4.6cm}}
    Input & ambient audio recording of a certain scene. \\
    \smallskip
    
    Output & class to which the scene belongs. \\
\end{tabular}
\medskip

The approach used not only challenges some of the fundamental mathematical techniques used so far in early experiments of the same trend, but also it introduces new scopes and new horizons for interesting results. The physics governing spectrograms has been optimized in the project along with exploring how it handles the intense requirements of the problem in hand. Major contributions and developments brought under the light through this project involve using better mathematical techniques and problem specific machine learning methods.

Improvised data analysis and data augmentation for audio datasets like frequency masking and random frequency-time stretching are used in the project and hence, are explained in this article. In the used methodology, audio transform principles were also explored, and indeed the insights gained were used constructively in the later stages of project. Using a deep learning approach is surely one of them.

Also, the potential scopes and upcoming research openings in both short and long term tunnel of time has been presented in this article. Although much of the results gained are domain specific as of now, but they are surely potent enough to produce novel solutions in various different domains of diverse backgrounds.
\medskip

\end{abstract}
\par\noindent\rule{7.1cm}{0.5pt}
\par\noindent\rule{7.1cm}{0.5pt}


\section{Introduction}
The automatic classification of environmental sound is a growing research field with multiple applications to large scale content-based multimedia indexing and retrieval.

While much of the literature and buzz on deep learning concerns computer vision and natural language processing (NLP), audio analysis — a field that includes automatic speech recognition (ASR), digital signal processing, and music classification, tagging, and generation — is a growing sub domain of deep learning applications.

To be precise, the sonic analysis of urban environments is the subject of increased interest, partly enabled by multimedia sensor networks, as well as by large quantities of online multimedia content depicting urban scenes. However, while there is a large body of research in related areas such as speech, music and \textit{bio-acoustics}, work on the analysis of urban acoustic environments is relatively scarce.

Significant research exposure is being provided to creative ideas which are potent enough to contribute novel solutions towards generic areas like noise-pollution control/reduction, and complicated and challenging areas like personalized location tagging with respect to noise levels/types etc.\\

\noindent This article is structured in a way where major \texttt{contributions} from this project is made towards the problem in hand is put next. Then the \texttt{backgrounds} section is kept followed by \texttt{Materials \& Methods} section, where the used procedural techniques are described in a detailed manner. Then \texttt{Discussions} are done followed by \texttt{Conclusions}. Finally the journal is halted with \texttt{Acknowledgements}.


\section{Contributions}

\noindent First of all, there has been many efforts in research and development sector to solve problems related to classification of land-use patterns. But majorly all the early work done in this context were entirely based on imagery datasets. This led to major public privacy breach issues and masses in general were not in favour of such experiments. This is based on the idea of using auditory datasets to solve the mentioned problems. Such approach is challenging indeed but it excludes the mentioned major issue related to privacy protocols.
\smallskip

Secondly, observing the mathematical techniques used in early and existing researches, it can be claimed that very less ideas were borrowed from other domains. Researchers usually used methods from \texttt{Linear Algebra} like matrix manipulation etc., which in general could not bear up the expected computational complexities. This is discussed in the upcoming sessions. This project works with \texttt{spectrograms} which is proven to fix the existing issues in computation.
\smallskip

Spectrograms can handle multiple parameters of audios at a time. This helps to fit the computational complexities within the expected levels. The physics governing spectrograms surely promises a better problem solving method for the problem in context.
\smallskip

Also, the lack of common vocabulary in datasets under use was a distinct problem for projects in the domain so far. In simple terms, for a certain acoustic scene like office, usually the quality of audio recorded from different locations like New York, Timbuktu and Wuhan differs with great extent. As a whole audio features like frequencies, average amplitude etc. also varies greatly. This issue is somewhat neutralized by the usage of spectrograms, as such audio features get evenly represented.
\smallskip

In this project, the data augmentation techniques used are in general a more specific option for the current problem. This further helps to analyzed data to fit better in the model. This is again discussed thoroughly in the Materials \& Methods section.
\smallskip

A pre-trained neural network is used to handle the \texttt{classification} problem component. Usage of such method specific neural network is currently gaining a lot of exposure in the mentioned domain. This contributes to the problem solving procedure by providing better results than other classical neural networks currently in use.
\smallskip

Also, on technical grounds, methods used to classify audio scenes ignored the temporal factors associated with audio datasets. The usage of problem specific convolutional neural network takes the \texttt{spatiotemporal} factor under consideration. Doing this evidently improves the quality of results.


\section{Background}

Motivated by various real-world applications, such as machine listening and surveillance, acoustic scene classification(ASC) has become a popular topic lately. Unlike in conventional sound processing, ASC deals with acoustics in daily life. Therefore, due to the complexity in environmental sound composition, conventional sound analysis schemes like ASR and music information retrieval, are not well suited to ASC.There are two channels of information that contribute to acoustic scene perception, namely, sound textures which present high temporal homogeneity nature of environmental sound and acoustic events which are superimposed on the background.


\section{Materials \& Methods}

\subsection{About the dataset}

This project makes use of auditory datasets from recordings made available as part of the DCASE (Detection and Classification of Acoustic Scenes and Events) 2019 challenge: \url{http://dcase.community/challenge2019/task-acoustic-scene-classification}

The \href{https://zenodo.org/record/2589280}{\texttt{dataset}} consists of 10-seconds audio segments from 10 acoustic scenes.
Each acoustic scene has 1440 segments (240 minutes of audio). The dataset contains in total 40 hours of audio.

\subsection{Data Analysis}

The meta-data of the audio files was extracted as depicted in the image: 
\begin{figure}[H]
  \centering
  \includegraphics[width=0.4\textwidth]{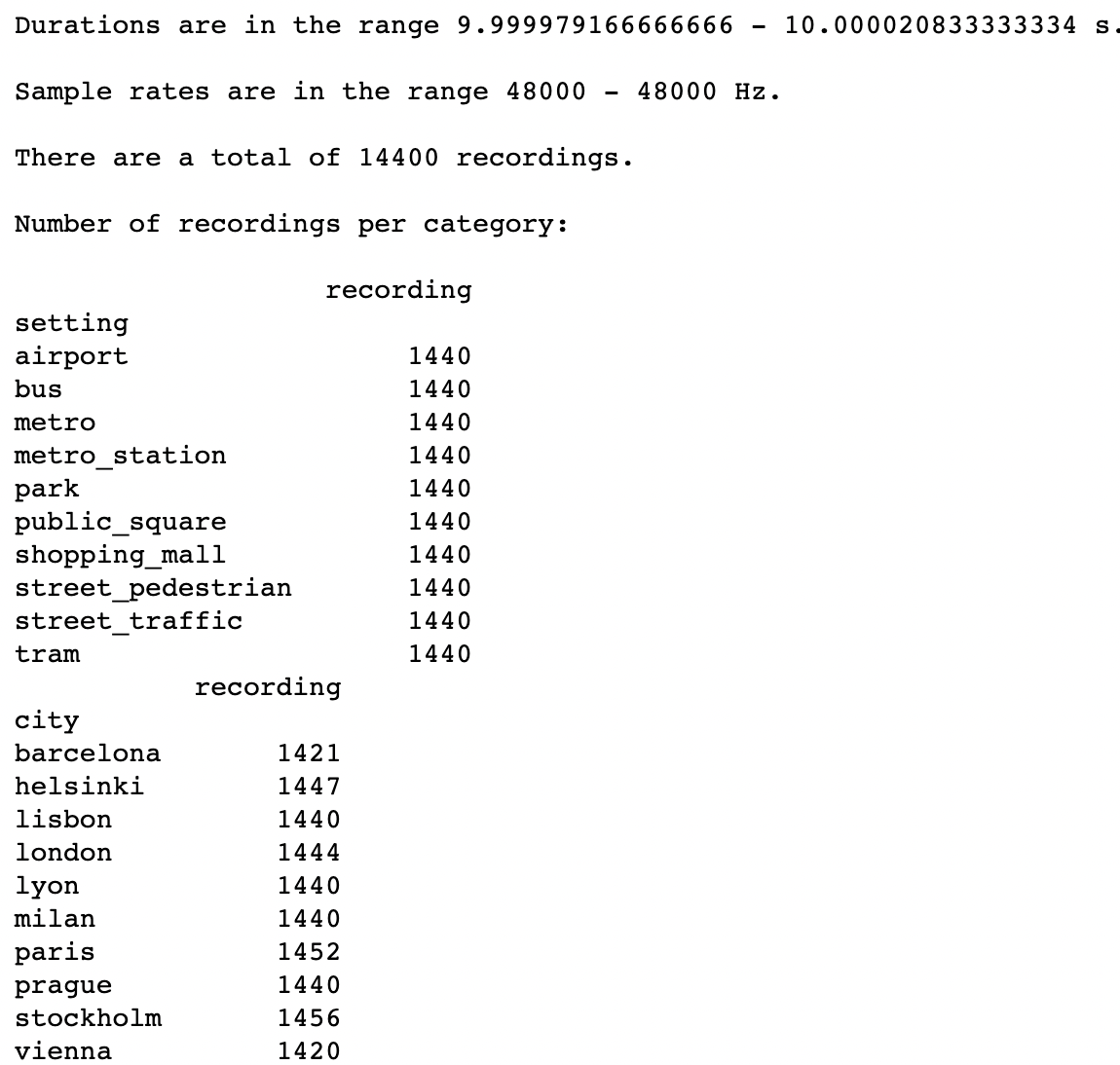}
  \caption{Metadata of the files}
\end{figure}

The primary approach of  the methodology was to convert audio files(.wav) to spectrograms(.png). In order to represent multiple factors associated with an audio file, which is also very essential for analysis and exploration, \textbf{Mel Spectrograms} were found to be the perfect agent.

\subsection{Spectrograms}

The Mel scale is based on what humans perceive as equal pitch differences. The Mel scale defines how the frequency axis is scaled in the spectrograms.\\

\noindent\textbf{1. }Evidently, the result of the scaling produced due to use of Mel Spectrograms densely distributed the low frequencies but in the same time it almost filtered out most of the high frequency tones (figure 2). In technical language we can say that the data represented was a bit skewed towards higher frequencies.\\
\begin{figure}[H]
  \centering
  \includegraphics[width=0.4\textwidth]{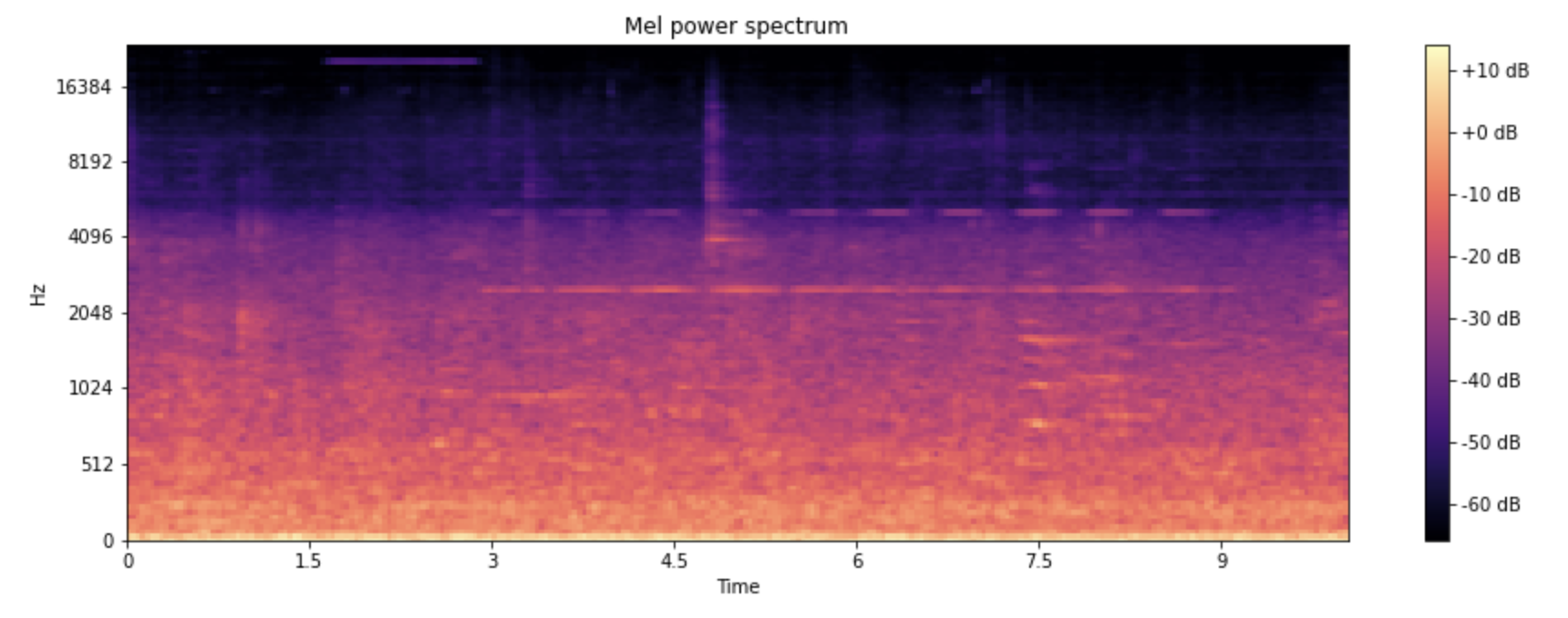}
  \caption{Mel Spectrogram.}
\end{figure}

\noindent\textbf{2. }To sort out this issue, the scaled spectrograms were further passed through a \textit{logarithmic re-scaler}, here in the form of \textbf{Log-Frequency Power Spectrogram}, also known as constant-Q power spectrogram. With the use of log-frequency spectrograms, the structure of low frequencies are gets well represented. Also the audio power is now more distributed over the frequency intervals (figure 3).\\
\begin{figure}[H]
  \centering
  \includegraphics[width=0.4\textwidth]{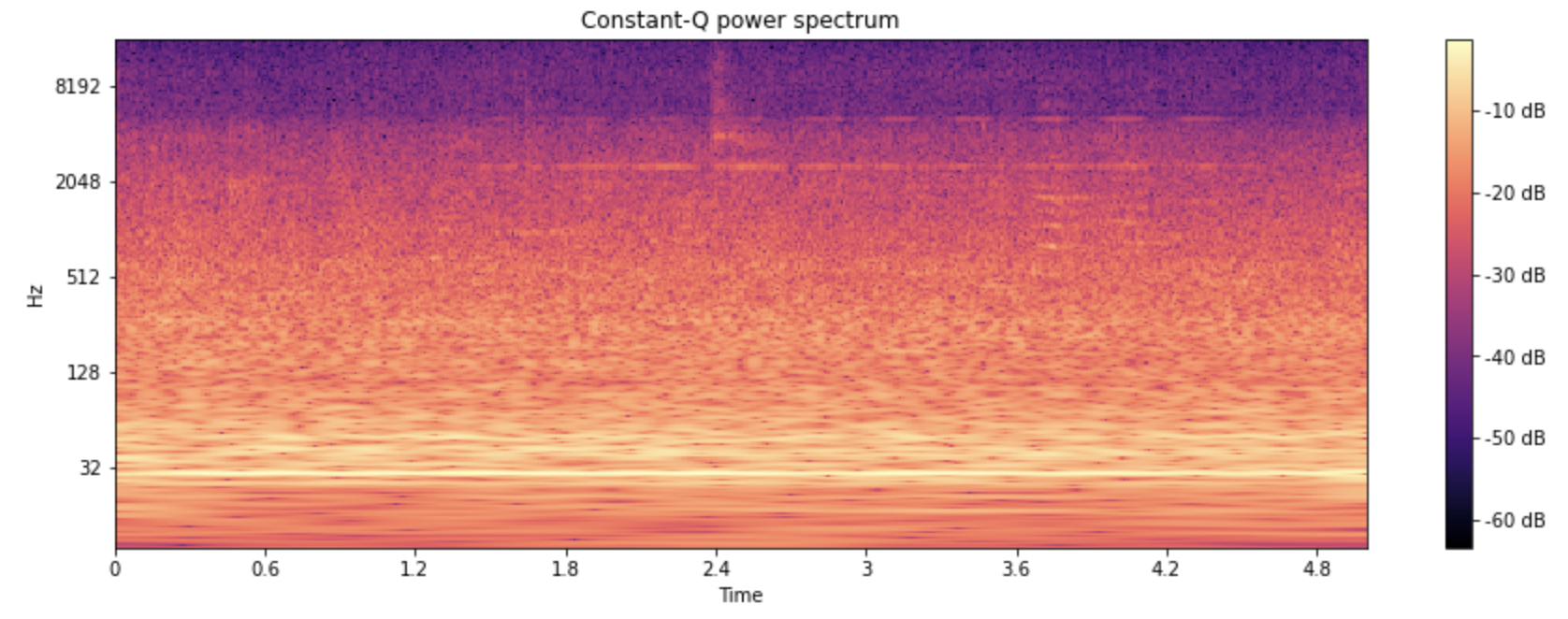}
  \caption{Log - Frequency Power Spectrogram (before pre-emphasis).}
\end{figure}

\noindent\textbf{3. }Yet high frequencies were still underrepresented. To sort out this issue, an emphasizing algorithm\cite{fayek2016}\cite{SpectroAlgo2} was used, which gradually toned the higher frequencies and fixed their representation. Broadly, the emphasizing algorithm filter mixes the signal with its first derivative. Mathematically,\begin{equation}y[n] = \frac{x[n] - (\alpha \cdot x[n-1])}{(1 - \alpha)}
\end{equation}
is summarised formula of the algorithm, where $\alpha$ is a constant (signal modulator). The code for the same is used in \textbf{Pre-Emphasis} part of the environment.\\
\begin{figure}[H]
  \centering
  \includegraphics[width=0.4\textwidth]{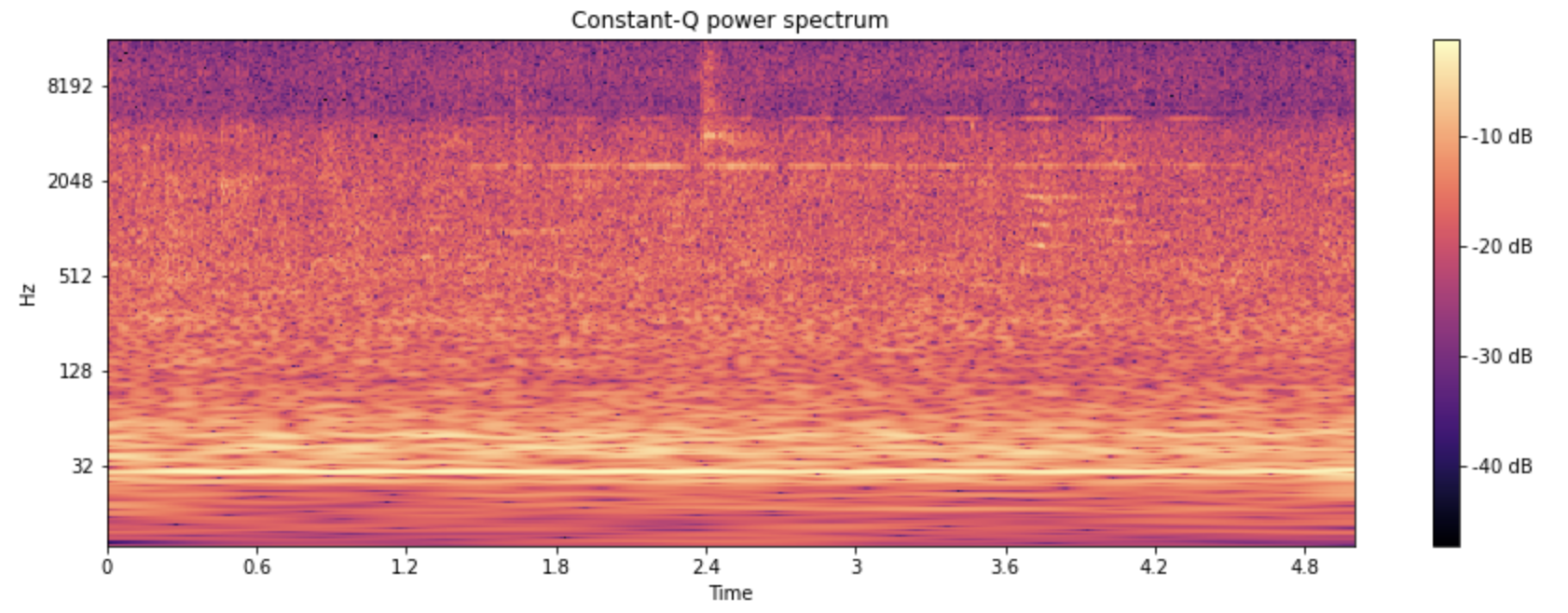}
  \caption{Log - Frequency Power Spectrogram (after pre-emphasis).}
\end{figure}

\noindent\textbf{4. }As a last block in the module, the resultant spectrograms are now converted to grey-scale images. This accounts for the simplification of the analysis complexity. These images act as an input for the model which shall be discussed further (figure 5).\\
\begin{figure}[H]
  \centering
  \includegraphics[width=0.4\textwidth]{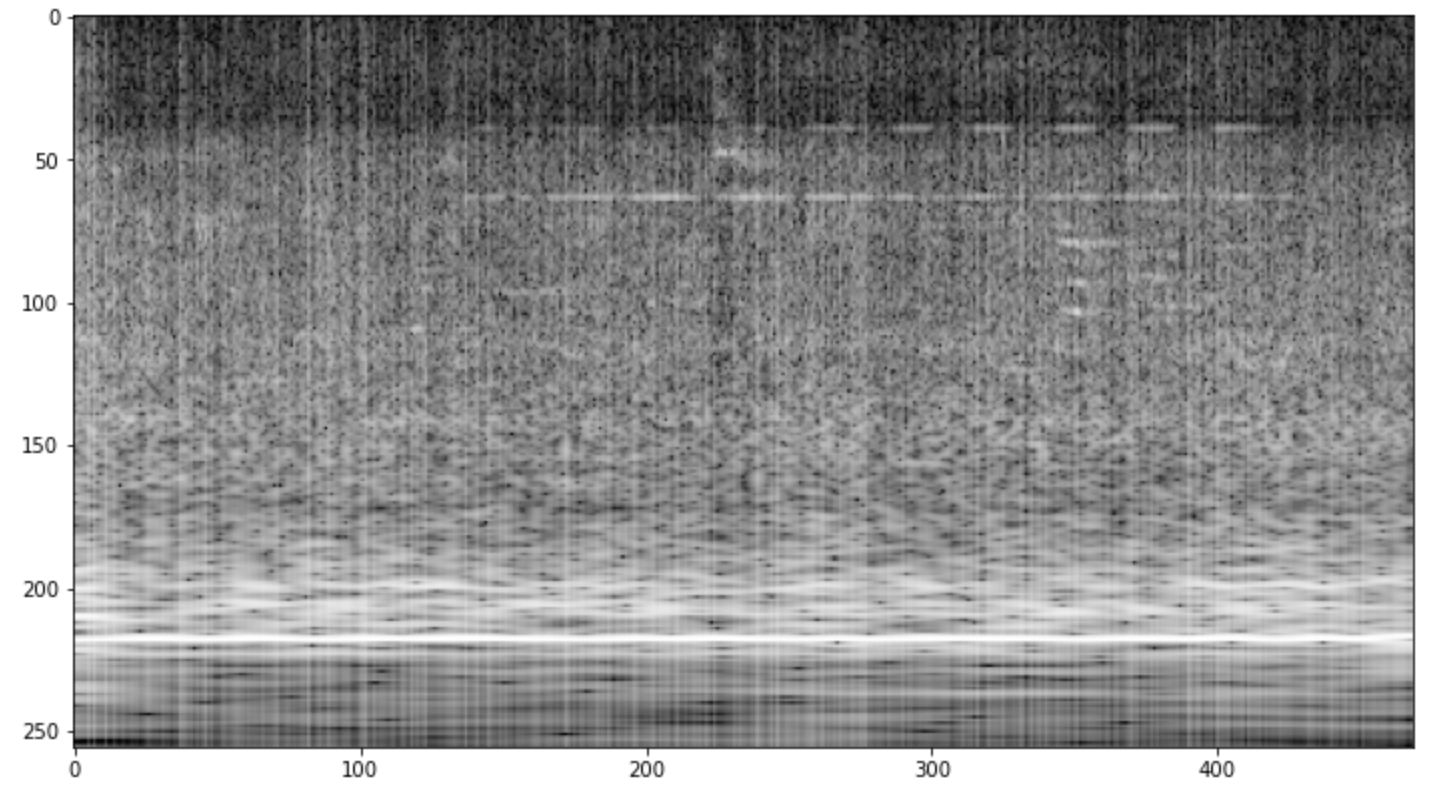}
  \caption{Final Spectrogram.}
\end{figure}

\noindent\texttt{\scriptsize A separate code file "Conversion\_to\_spectrograms.py" was
created which enables a single terminal command to convert 
all audio files into spectrograms.}

\subsection{Modelling}

Now the problem becomes an image classification one rather than an audio classification problem. So, these images need to be fed to the model viz. a convoluted neural network model(CNN). But there is a caveat here: the data requirements of a neural network model is quite high as opposed to 1440 per class in this present context (figure 1). This is addressed by augmenting the data.\\

\par\noindent\rule{7.1cm}{0.5pt}\\
\noindent\textbf{Data Augmentation}
\par\noindent\rule{7.1cm}{0.5pt}\\

\noindent Just like with images, there are several techniques to augment audio data as well. This augmentation can be done both on the raw audio before producing the spectrogram, or on the generated spectrogram. Augmenting the spectrogram usually produces better results.

\subsubsection{Raw Audio Augmentation}

\noindent\texttt{Time Stretch} : Stretching the time axis according to the sample size.The sound is randomly slowed down or speed up.
Comparing figure 6 and figure 7, the difference in the graduations on the axes can be noticed.

\begin{figure}[H]
  \centering
  \includegraphics[width=0.4\textwidth]{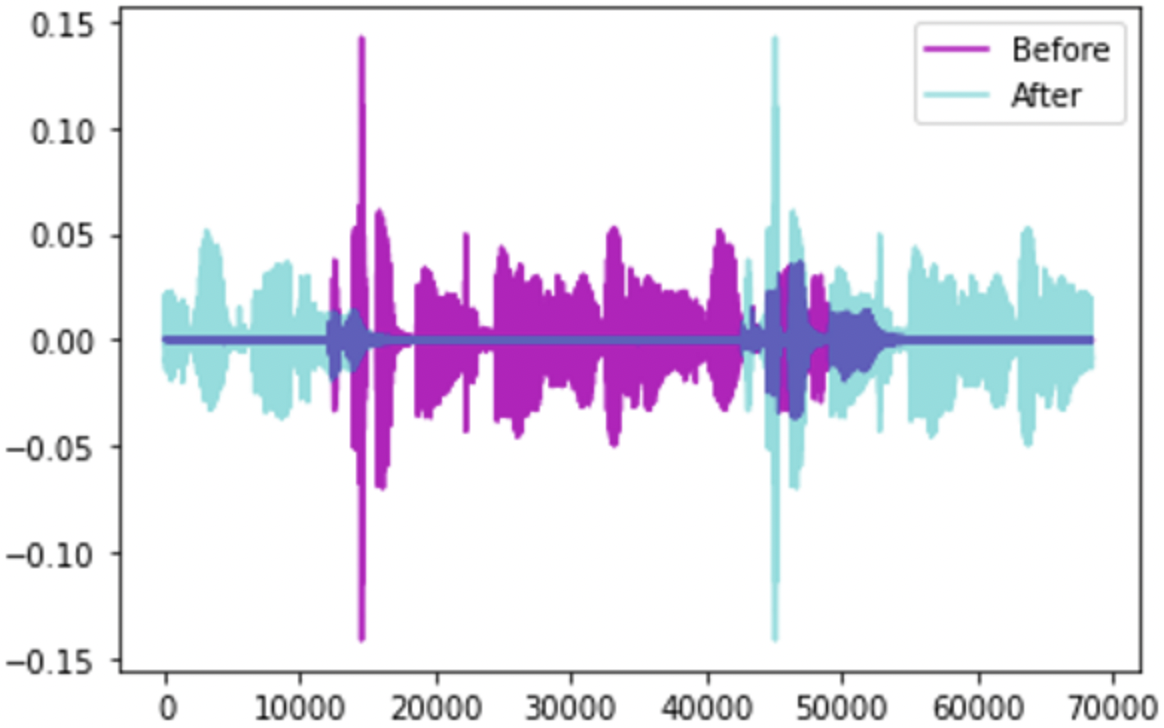}
  \caption{Time Stretch.}
\end{figure}

\begin{figure}[H]
  \centering
  \includegraphics[width=0.4\textwidth]{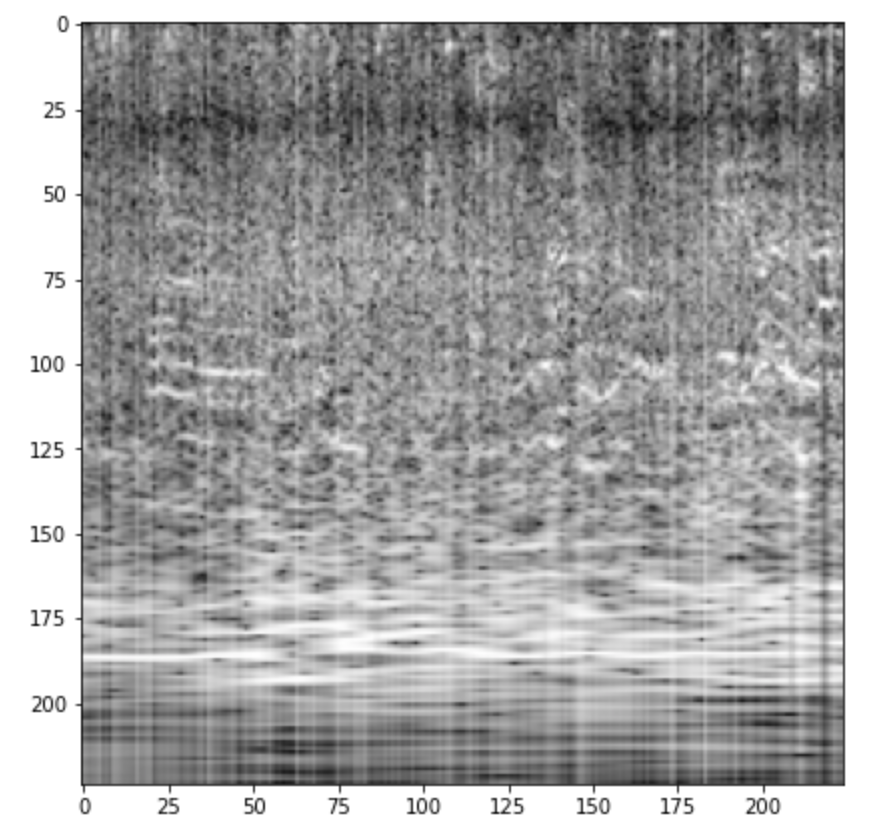}
  \caption{Image after Random Audio Transform.}
\end{figure}

\subsubsection{Spectrogram Augmentation}

\noindent\texttt{Frequency mask} — Randomly masking out a range of consecutive frequencies by adding horizontal bars on the spectrogram (figure 8).

\begin{figure}[H]
  \centering
  \includegraphics[width=0.4\textwidth]{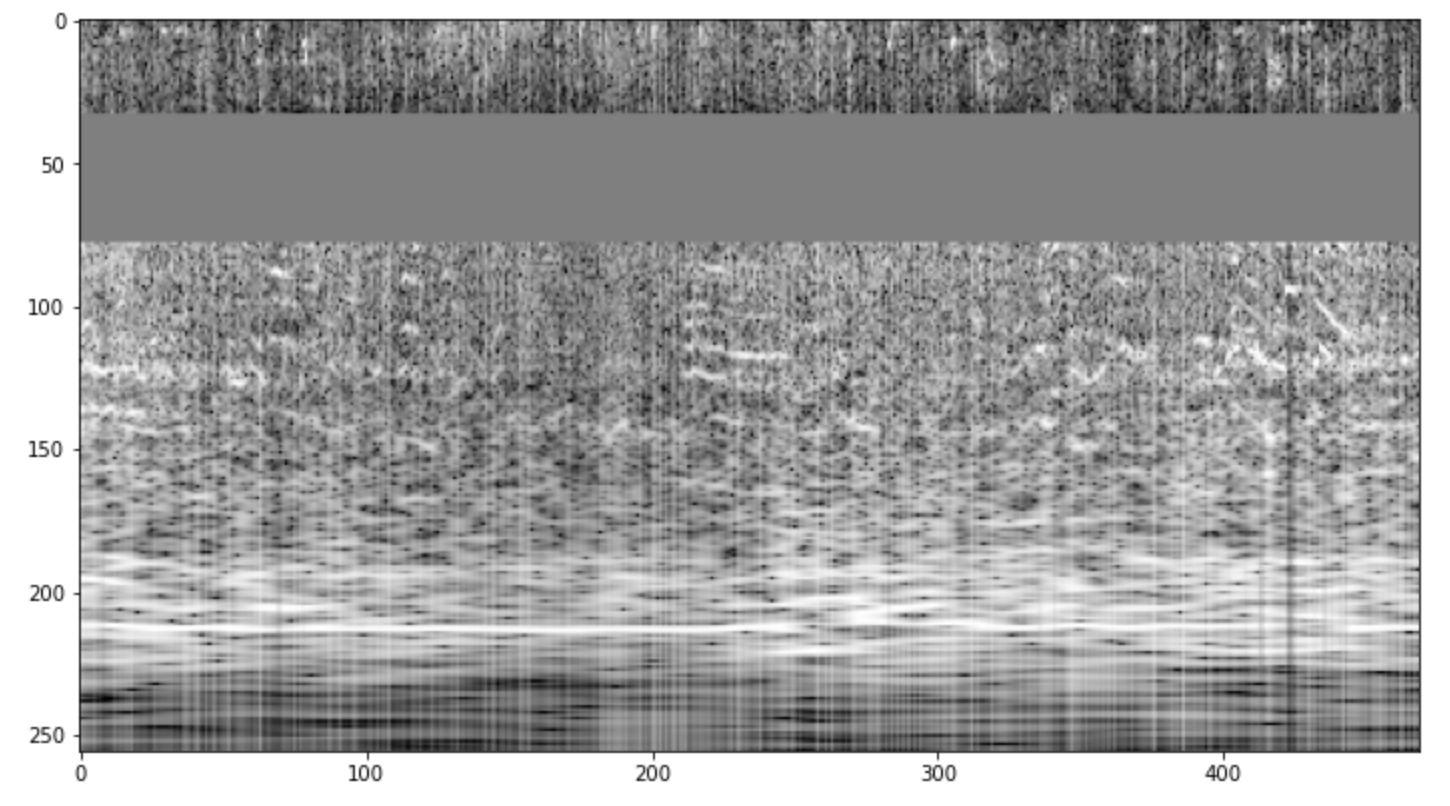}
  \caption{Image after frequency masking.}
\end{figure}

\vspace*{\fill}
\noindent\texttt{\scriptsize Before training the data (spectrograms), it was splitted
into the train, validation and test datasets. This was 
done using the script python file "Splitting\_datasets.py".}
\newpage

\par\noindent\rule{7.1cm}{0.5pt}\\
\noindent\textbf{Neural Network Training}
\par\noindent\rule{7.1cm}{0.5pt}\\

\noindent Using a pre-trained network as a starting point simplifies finding solutions. One such example for image recognition is \textit{VGG16} (trained on the ImageNet dataset) delivered with the \textit{torchvision} package\cite{jayesh1}.
In this project, a pre-trained VGG16 model with custom classifiers was employed.

\begin{figure}[H]
  \centering
  \includegraphics[width=0.4\textwidth]{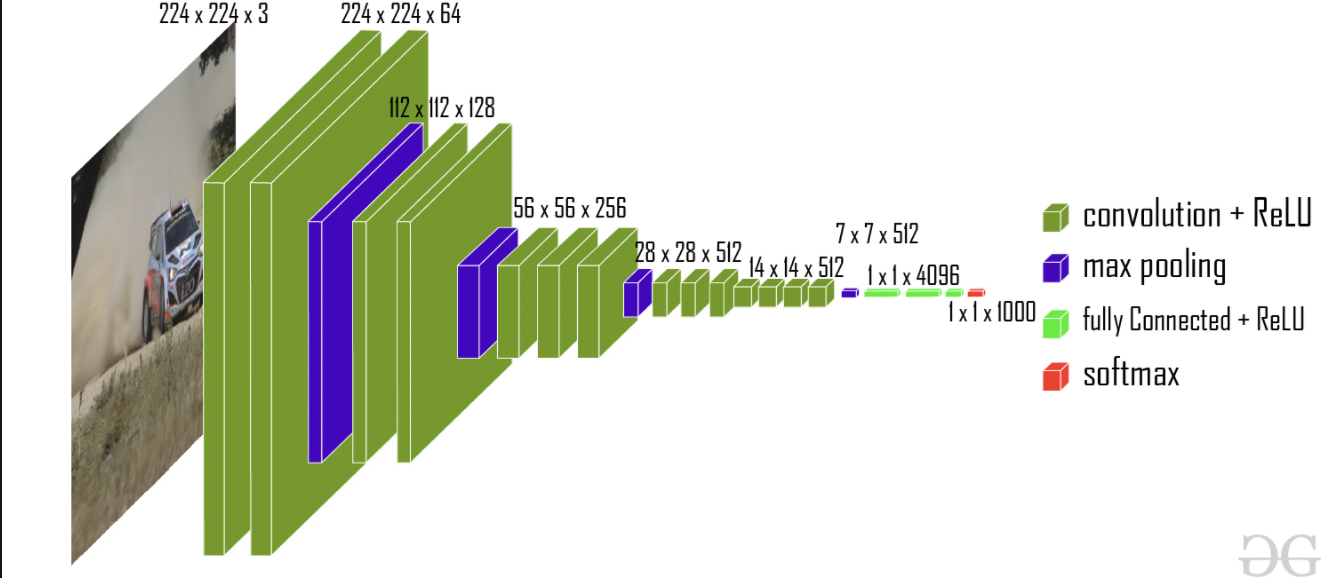}
  \caption{VGG16 model with additional layers.}
\end{figure}

\subsubsection{Create Transforms}

Several hyper-parameters like \textit{octaves}\href{https://en.wikipedia.org/wiki/Octave}{\textbf{*}}, number of \textit{bins per octave} were defined. The data was normalized using a default normalization vector for pre-trained VGG16. The train and test transforms were applied after applying the augmentation functions.\\ 

\subsubsection{Define Custom Data Loaders}

A function was defined that fetches the individual data items and packages them into a batch of data. In this project, the data was loaded as batches of 32.\\
\begin{figure}[H]
  \centering
  \includegraphics[width=0.4\textwidth]{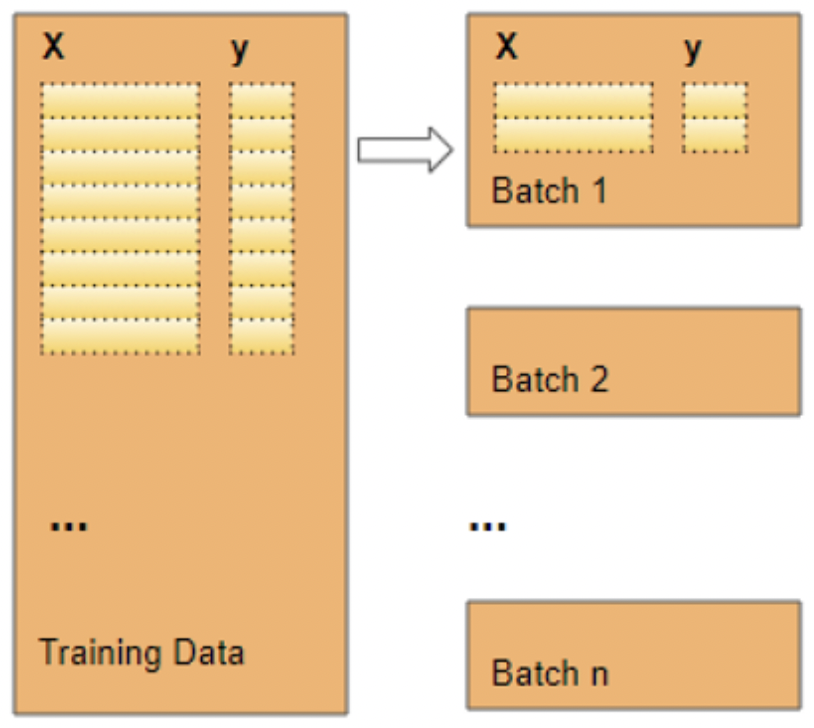}
  \caption{Data gets fed in batches of 32}
\end{figure}

\subsubsection{Creating Model}

Two fully connected layers with ReLU activation and another fully connected layer with Softmax activation were created as classifiers layers. A function for loading the model, with pre-trained VGG16 and the classifiers was loaded.
\begin{figure}[H]
  \centering
  \includegraphics[width=0.4\textwidth]{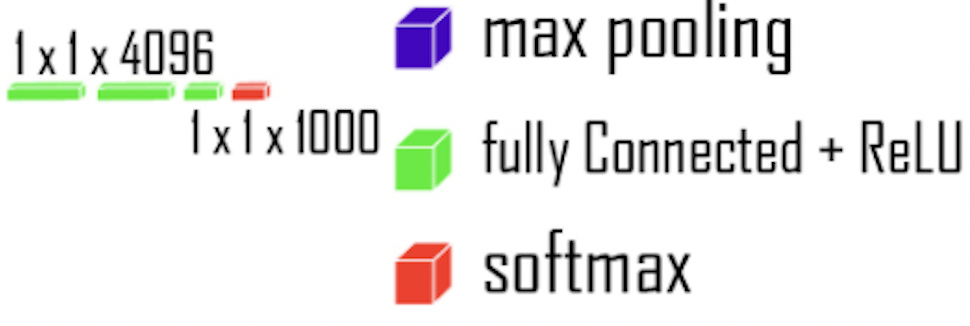}
  \caption{Custom Classifiers}
\end{figure}

\subsubsection{Training}

After defining the functions for measuring accuracy and saving weights; the training loop was run. An accuracy of 68\% was achieved in the 24\textsuperscript{th} epoch. After running till the 34\textsuperscript{th} epoch and the accuracy not improving; the training loop terminated. Hence, an accuracy of around 68\% on the validation set was obtained.
\begin{figure}[H]
  \centering
  \includegraphics[width=0.4\textwidth]{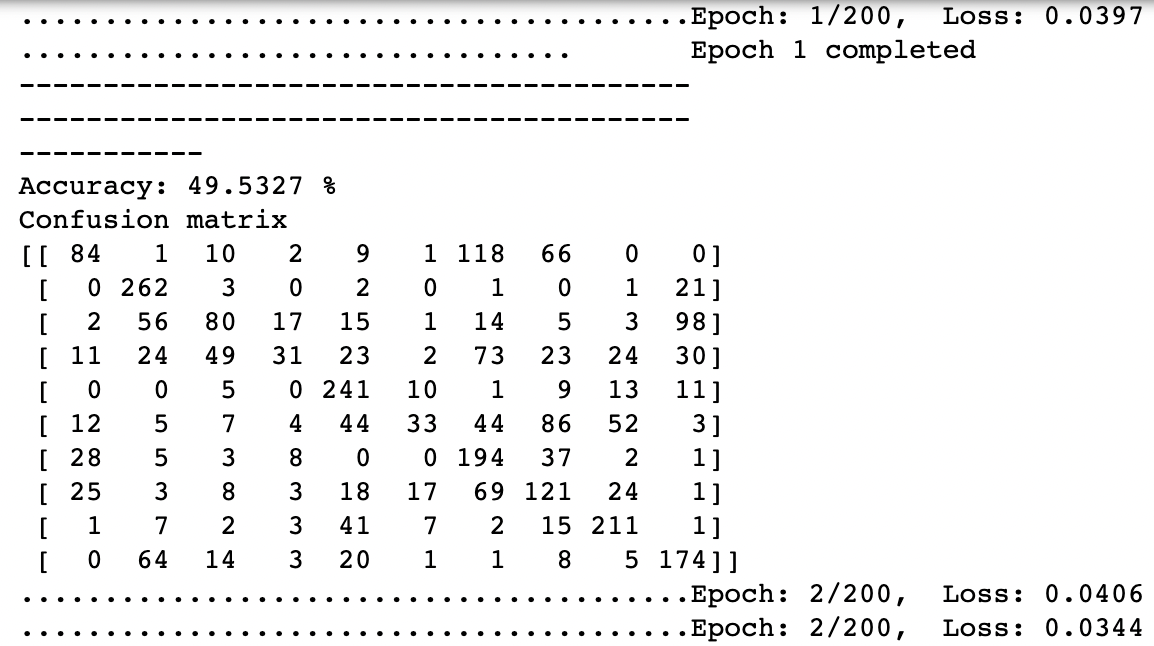}
  \caption{Epoch 1}
\end{figure}
\begin{figure}[H]
  \centering
  \includegraphics[width=0.4\textwidth]{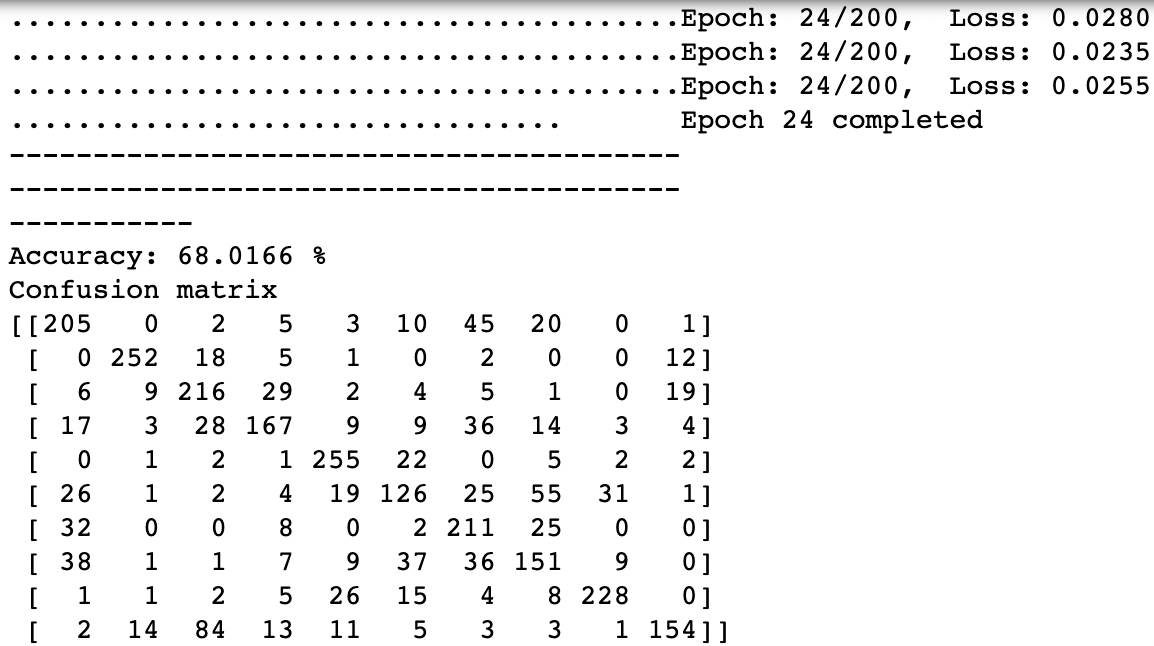}
  \caption{Epoch 24}
\end{figure}
\begin{figure}[H]
  \centering
  \includegraphics[width=0.4\textwidth]{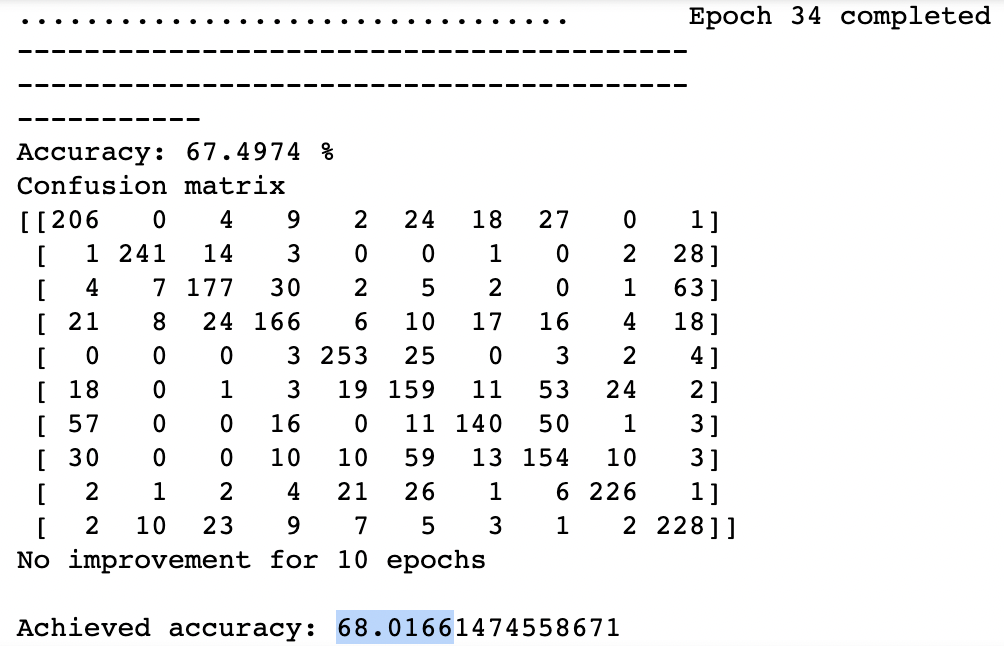}
  \caption{Training end at Epoch 34}
\end{figure}

\par\noindent\rule{7.1cm}{0.5pt}\\
\noindent\textbf{Evaluation}
\par\noindent\rule{7.1cm}{0.5pt}\\

\noindent Upon testing the model on the testing dataset, an accuracy of around 67.6\% was achieved. The confusion matrix is also depicted in the following image:
\begin{figure}[H]
  \centering
  \includegraphics[width=0.4\textwidth]{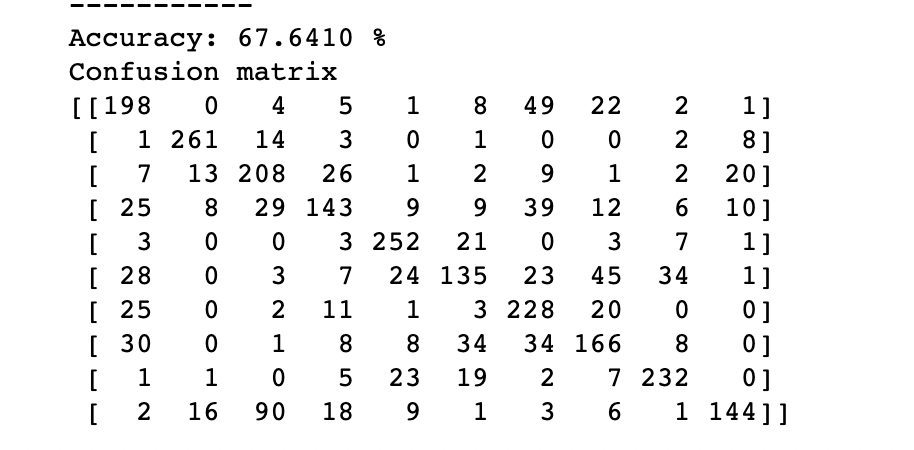}
  \caption{Confusion Matrix}
\end{figure}
\begin{figure}[H]
  \centering
  \includegraphics[width=0.4\textwidth]{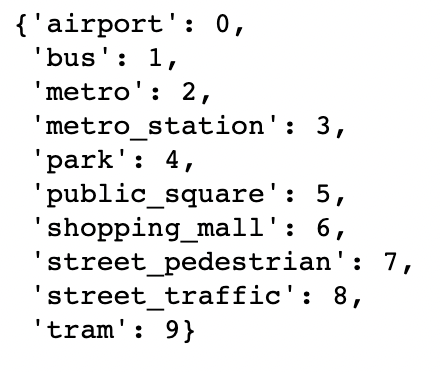}
  \caption{Class-indices for reading the matrix}
\end{figure}


\section{Results}

The model is now trained and evaluated. Now the aim is to try and predict something! 
We input an audio file to our model. The results are depicted in figure 17:
\begin{figure}[H]
  \centering
  \includegraphics[width=0.4\textwidth]{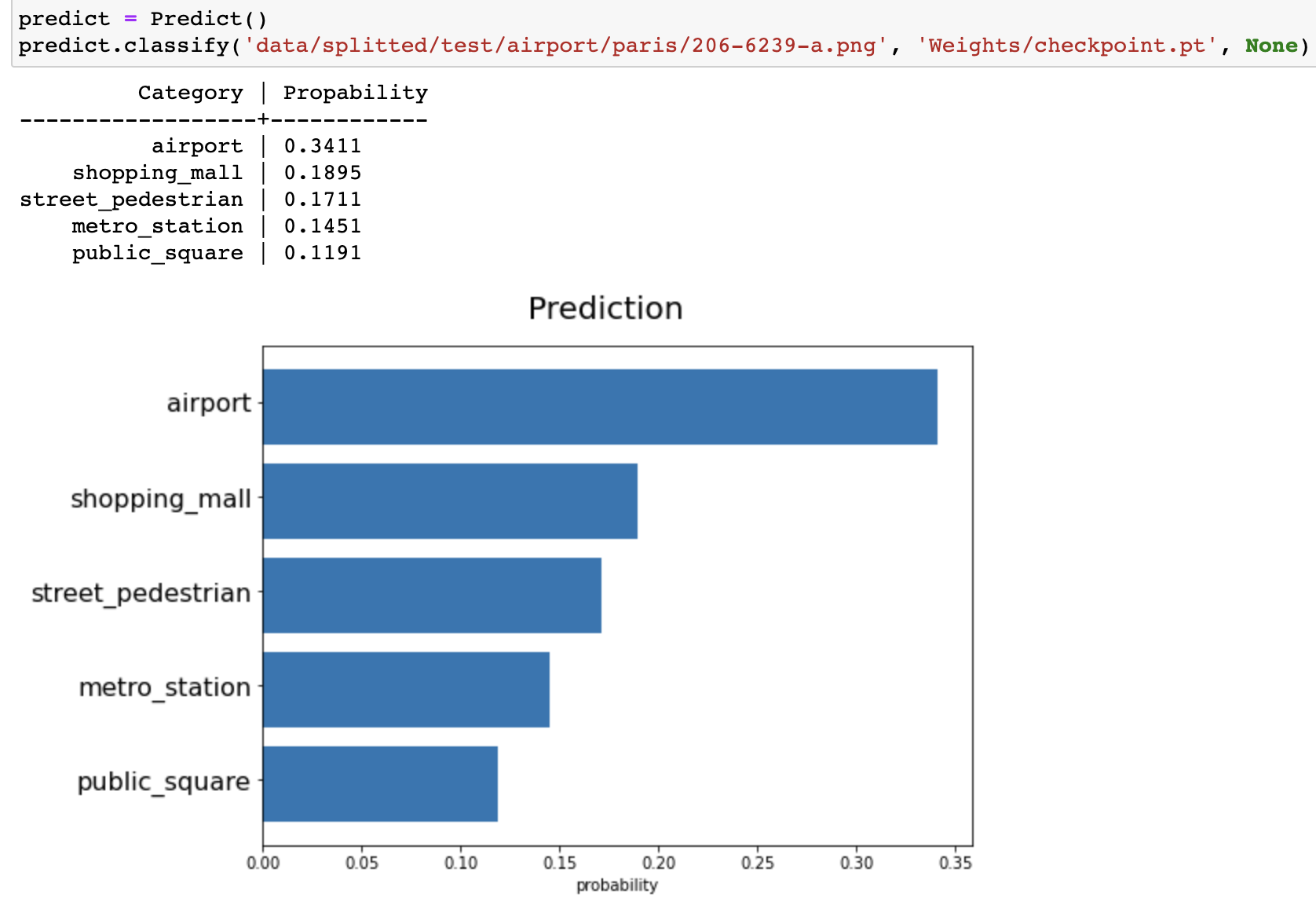}
  \caption{Probability graph for prediction}
\end{figure}


\section{Discussion}

\subsection{Classical techniques refined}

\par\noindent\rule{7.1cm}{0.5pt}\\
\noindent\textbf{Pre-processing}
\par\noindent\rule{7.1cm}{0.5pt}\\

\noindent When it comes to data pre-processing, several techniques based on some core mathematical concepts were under consideration. A method that was being explored in some early experiments was regarding the usage of \href{https://en.wikipedia.org/wiki/Array}{\texttt{arrays}}.\\

\noindent\textbf{1. }In such methods, only one factor of audio files in general is taken at a time, for example \textit{frequency}, and then the audio file is converted into an array using sampling techniques\href{https://www.kdnuggets.com/2020/02/audio-data-analysis-deep-learning-python-part-1.html}{\textbf{*}}.\\ 
\begin{figure}[H]
  \centering
  \includegraphics[width=0.4\textwidth]{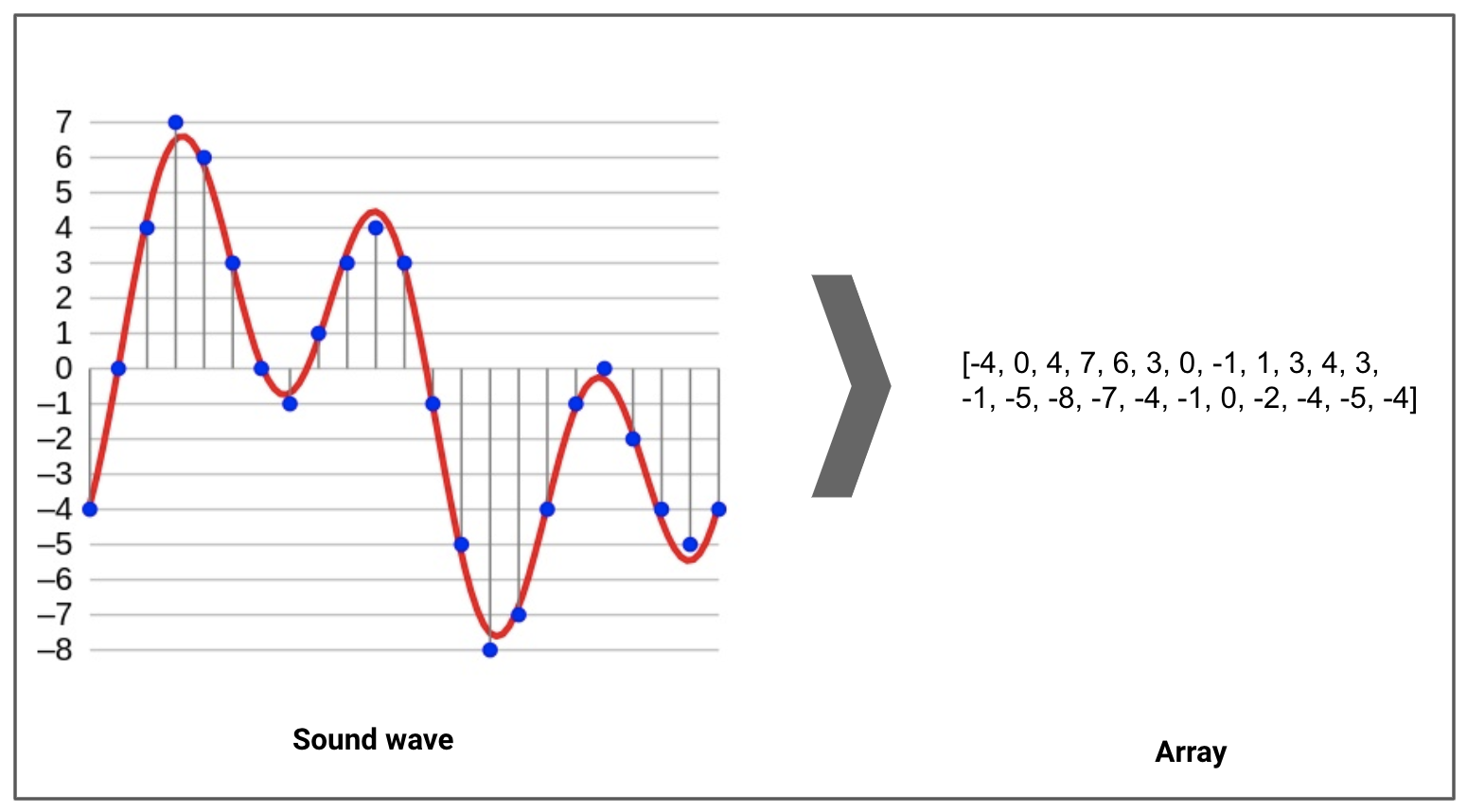}
  \caption{Audio to array.}
\end{figure}

\noindent\textbf{2. }After getting arrays of different properties of the audio files, the arrays are then used together to frame out a multidimensional array or matrix. This matrix is the passed through several mathematical and computational techniques based on \textbf{linear algebra} for analysis\cite{mularr}.\\
\begin{figure}[H]
  \centering
  \includegraphics[width=0.4\textwidth]{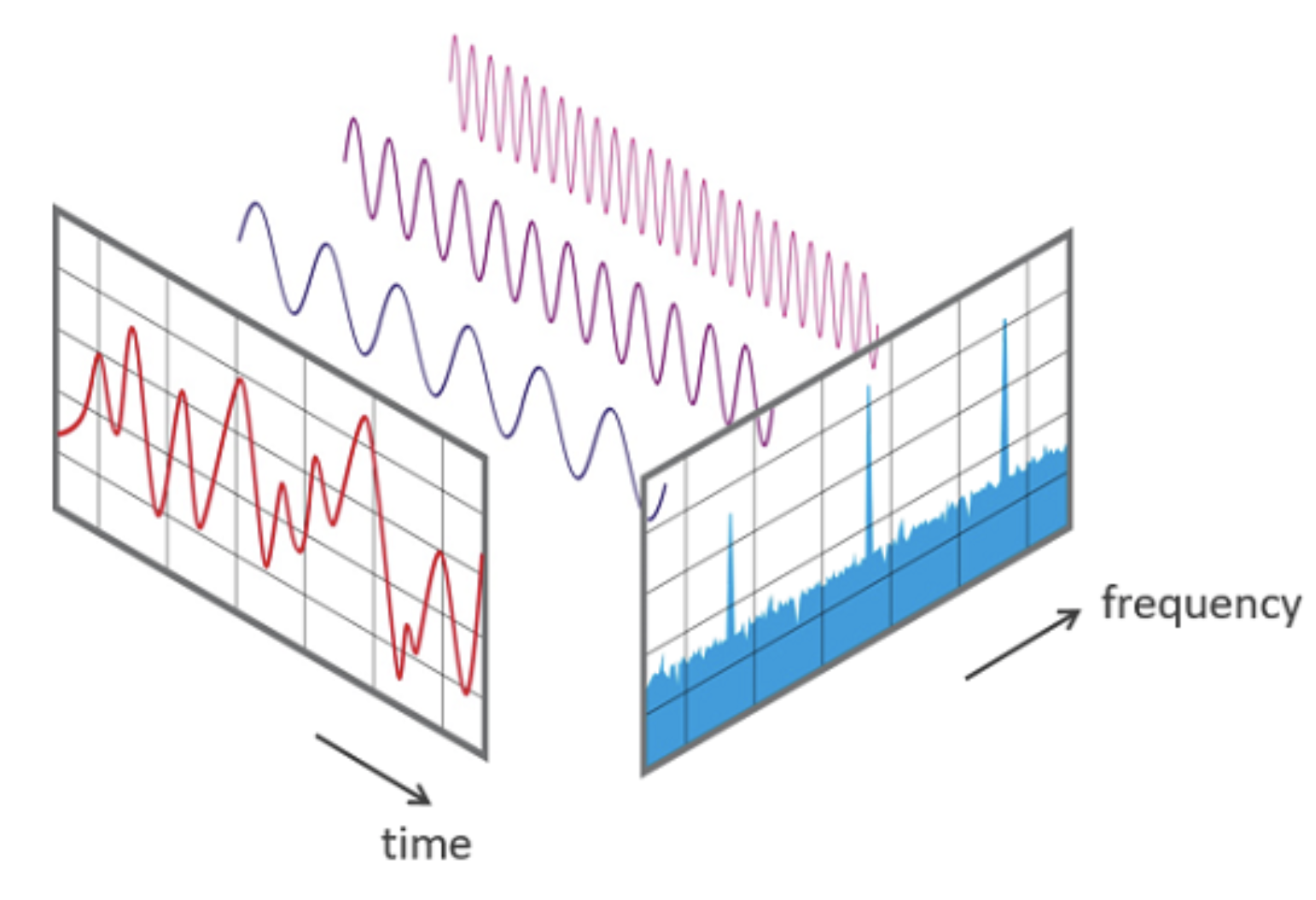}
  \caption{Audio to multidimensional array.}
\end{figure}

But this methods could not meet the expectations required because often it could not bear the computational complexities needed for solving the problem in hand. In such case spectrograms were found better according to the requirements.\\

\par\noindent\rule{7.1cm}{0.5pt}\\
\noindent\textbf{Neural Network}
\par\noindent\rule{7.1cm}{0.5pt}\\

\noindent ASC algorithms mostly use CNN based network architectures since they usually provide a summarizing classification of longer acoustic scene excerpts.\\

\noindent A pre-trained VGG16 proved useful because this model is trained on a huge dataset. Maybe, this is the reason why it proved useful for audio classification problems which have less amount of datasets.\\

\noindent In contrast, AED (Acoustic Event Detection) algorithms commonly use convolutional recurrent neural networks (\textit{CRNN}) as they focus on a precise detection of sound events. This architecture combines CNN as the front-end for representation learning and a recurrent layer for temporal modeling. Unfortunately, due to various constraints this could not be tried and implemented by the team. 

\subsection{Limitations faced}

\noindent First of all, the usage of large datasets improved the quality of probabilistic predictions as a whole. But simultaneously, on the other hand it also increased the need of high end hardware to load the required computational needs. The problem faced was specific to the lack of high-end hardware tools like better GPU etc. As a result of which, the process of repeated training of neural network and running the overall coding setup in general took too much time. Secondary methods like usage of Google Collab was used at last.\\

A problem faced which technically cannot be refined further is the memory requirements involved while storing the audio files and altogether with the converted spectrograms. It is almost certain that if the size of the input dataset is increased for refined results then the intensity of this problem will increase likewise.\\

Also, while executing the project in successive phases, a need of using better data augmentation methods was felt. It was observed that there was a scope of improvement.


\section*{Conclusions}
If the project is further approached, then surely some major refinements would be tried for. In general, the discussed short term and long term proposals and ideas are an indirect implication of some difficulties faced during the conduct of the project. The team as a whole felt that such hindrances should be solved for the betterment of the project in further stages. Likewise some suggestions would be as follows:
\begin{itemize}
    \item[1.] Optimizing the parameters used for data augmentation and the transforms involved within is one such proposal. Most likely, as backed up by the exposure gained through this project, it would be beneficial to try grid search for this task\cite{aug++},\cite{unknown}.
    
    \item[2.] Moving ahead, many other data augmentation techniques like usage of proportional/strict filtering of ambient noise should also be explored\cite{noice}.
    
    \item[3.] A simple but indeed a challenging task would be to design specific neural networks for acoustic scene classification. Even the prospect of designing a neural network with increased specificity might tend to open further interesting research gateways\cite{mult-cnn},\cite{cnn7}.
\end{itemize}

Talking about further strides, many active researches are already under progress which promise interesting contribution towards the domain of environmental science. To be one step specific, we can point out advancements done in the areas like bio-acoustic for animal sound classification\cite{anml},\cite{anml2}.

Creatively, results obtained at the end of this project might end up helping defence technologies. Domains like security surveillance of remote and inaccessible places and areas need further refinements in their existing technology. Suspected personalities and targets can be tagged through ambient noise and sounds. This seems to be a very promising possibility as \textbf{acoustic scene classification} methods are already a key in such scenarios.  


\section*{Acknowledgements}

This project is a result of valuable support from Dr. Vaibhav Kumar of Data Science and Engineering Dept., Indian Institute of Science Education and Research, Bhopal. His insightful suggestions and creative comments as a mentor was a constant source of motivation throughout this journey.

The deep learning approaches used in the project is an implied result of thorough literature explorations. Ideas discussed in the journal issued by \textbf{Semantic Music Technologies}\cite{app10062020} on deep learning based methods was very insightful.

Many research sources proved to be of great value in order to shape the structure of the project. This IEEE paper\cite{7078982} was helpful in providing some structure.

Journals\cite{jung2021dcasenet},\cite{koutini2019receptivefieldregularized} provided essential theories and creative ideas, necessary for giving a directional thrust to the solution. Thanks of gratitude to the authors.

This \href{https://towardsdatascience.com/audio-deep-learning-made-simple-sound-classification-step-by-step-cebc936bbe5}{\texttt{blog}} published on \textbf{Towards Data Science} series was also a good read to start.

And last but not the least, in very early stages of the project execution, it had become very important to get a hold on what exactly the road-map of the project should be. In such tough times the following projects helped the team to form a proper road-map for this project.
\begin{itemize}
    \item Sound Classification using Spectrogram Images | \href{https://www.kaggle.com/devilsknight/sound-classification-using-spectrogram-images/notebook}{\texttt{kaggle.com}}
    \item Audio-Scene-Classification | \href{https://github.com/anujdutt9/Audio-Scene-Classification}{\texttt{github.com}}
\end{itemize}

\bibliography{bibliography}

\begin{thebibliography}{10}

\bibitem{fayek2016}
Haytham~M. Fayek.
\newblock
  \href{https://haythamfayek.com/2016/04/21/speech-processing-for-machine-learning.html}{Speech
  Processing for Machine Learning: Filter banks, Mel-Frequency Cepstral
  Coefficients (MFCCs) and What's In-Between}, 2016.

\bibitem{SpectroAlgo2}
Abdel~Rahman Mohamed.
\newblock
  \href{https://tspace.library.utoronto.ca/bitstream/1807/44123/1/Mohamed_Abdel-rahman_201406_PhD_thesis.pdf}{Deep
  Neural Network acoustic models for ASR}.
\newblock {\em University of Toronto}, 2014.

\bibitem{jayesh1}
Karen Simonyan and Andrew Zisserman.
\newblock \href{https://arxiv.org/abs/1409.1556.pdf}{Very Deep Convolutional
  Networks for Large-Scale Image Recognition}, 2014.

\bibitem{mularr}
Ashu M.~G. Solo.
\newblock
  \href{http://www.iaeng.org/publication/WCE2010/WCE2010_pp1829-1833.pdf}{Multidimensional
  Matrix Mathematics: Multidimensional Matrix Equality, Addition, Subtraction,
  and Multiplication, Part 2 of 6}.
\newblock {\em Proceedings of the World Congress on Engineering 2010}, III,
  2010.

\bibitem{aug++}
Helin Wang, Yuexian Zou, and Wenwu Wang.
\newblock
  \href{https://www.isca-speech.org/archive/pdfs/interspeech_2021/wang21d_interspeech.pdf}{SpecAugment++:
  A Hidden Space Data Augmentation Method for Acoustic Scene Classification}.
\newblock {\em INTERSPEECH 2021}, 2021.

\bibitem{unknown}
Helen Bear, Veronica Morfi, and Emmanouil Benetos.
\newblock An evaluation of data augmentation methods for sound scene
  geotagging, 10 2021.

\bibitem{noice}
Yongju Choi, Othmane Atif, Jonguk Lee, Daihee Park, and Yongwha Chung.
\newblock Noise-robust sound-event classification system with texture analysis.
\newblock {\em Symmetry}, 10:402, 09 2018.

\bibitem{mult-cnn}
Dawei Feng, Haibo Mi, Boqing Zhu, Kele Xu, Sheuwen Liu, Dezhi Wang, Lilun
  Zhang, and Hengxing Cai.
\newblock \href{https://arxiv.org/pdf/1805.07319.pdf}{Mixup-Based Acoustic
  Scene Classification Using Multi-Channel Convolutional Neural Network}, 2018.

\bibitem{cnn7}
Thomas Lidy and Alexander Schindler.
\newblock
  \href{http://www.ifs.tuwien.ac.at/~schindler/pubs/DCASE2016b.pdf}{CQT-based
  convolutional neural networks for audio scene classification and domestic
  audio tagging}, 2016.

\bibitem{anml}
Loris Nanni, Gianluca Maguolo, and Michelangelo Paci.
\newblock \href{https://arxiv.org/pdf/1912.07756.pdf}{Data augmentation
  approaches for improving animal audio classification}, 2018.

\bibitem{anml2}
Loris Nanni, Gianluca Maguolo, and Michelangelo Paci.
\newblock Data augmentation approaches for improving animal audio
  classification.
\newblock {\em Ecological Informatics}, 57:101084, 03 2020.

\bibitem{app10062020}
Jakob Abeßer.
\newblock \href{https://www.mdpi.com/2076-3417/10/6/2020/htm}{A Review of Deep
  Learning Based Methods for Acoustic Scene Classification}.
\newblock {\em Applied Sciences}, 10(6), 2020.

\bibitem{7078982}
Daniele Barchiesi, Dimitrios Giannoulis, Dan Stowell, and Mark~D. Plumbley.
\newblock
  \href{https://ieeexplore.ieee.org/stamp/stamp.jsp?tp=&arnumber=7078982}{Acoustic
  Scene Classification: Classifying environments from the sounds they produce}.
\newblock {\em IEEE Signal Processing Magazine}, 32(3):16--34, 2015.

\bibitem{jung2021dcasenet}
Jee weon Jung, Hye jin Shim, Ju~ho~Kim, and Ha-Jin Yu.
\newblock \href{https://arxiv.org/pdf/2009.09642v1.pdf}{DcaseNet: An integrated
  pretrained deep neural network for detecting and classifying acoustic scenes
  and events}, 2021.

\bibitem{koutini2019receptivefieldregularized}
Khaled Koutini, Hamid Eghbal-zadeh, and Gerhard Widmer.
\newblock
  \href{https://arxiv.org/pdf/1909.02859v1.pdf}{Receptive-field-regularized CNN
  variants for acoustic scene classification}, 2019.

\end{thebibliography}

\vspace*{\fill}
\par\noindent\rule{\textwidth}{0.4pt}

\noindent\texttt{Reach us by - }\\
\noindent\texttt{Jayesh Kumpawat $\cdots\cdot$}  jayesh19@iiserb.ac.in\\
\noindent\texttt{Shubhajit Dey $\cdots$}  shubhajit19@iiserb.ac.in

\end{document}